\documentstyle[aps,prd,preprint]{revtex}

\begin{document}
\draft


\title{A geometrical action for dilaton gravity}

\author{Alberto Saa}

\address{Departamento de F\'\i sica de Part\'\i culas,\\ 
Universidade de Santiago de Compostela,\\
E-15706, Santiago de Compostela, Spain }


\maketitle

\begin{abstract}
We study the gravitational interaction involving the dilaton
and the anti-symmetrical $B_{\mu\nu}$ fields that arises in the 
low-energy limit of string theory. It is shown that such
interaction can be derived from a geometrical action principle,
with the scalar of curvature of a non-Riemannian, but {\em metric-compatible},
connection as the lagrangian, and with a non-parallel volume-element.
This action is contrasted with the recently proposed  geometrical action
for the 4-dimensional axi-dilaton gravity. 
\end{abstract}
\pacs{PACS Numbers: 0450, 1125}

Recently, Derely and Tucker\cite{DT} have shown that the 4-dimensional
axi-dilaton gravity equations can be obtained from a geometrical action
principle, where the lagrangian is given by 
the scalar of curvature of a non-Riemannian
connection. 
Besides of be an elegant way of
understanding
 the axi-dilaton gravity, such result provides us with a geometrical
interpretation for the massless states of the closed bosonic string. Since
the equations for
 all the massless states of the closed string (namely the dilaton, the
graviton, and the anti-symmetrical $B_{\mu\nu}$ field) 
arise in the same footing from the
requirement of conformal invariance up to one-loop order of the
closed string in background fields\cite{CFMP,GSW},
such common interpretation in geometrical terms to these states is
natural and pertinent.
The seeds of this geometrical interpretation  can be encountered in
the middle 70's works of Scherk and Schwarz\cite{SS}.
In the model presented in \cite{DT},
the torsion tensor is proportional to the anti-symmetrical
field $H_{\alpha\beta\gamma} = \partial_\alpha B_{\beta\gamma} +
\partial_\gamma B_{\alpha\beta} +\partial_\beta B_{\gamma\alpha}$,
in agreement with previous works on sigma models\cite{BCZ}, and the
non-metricity to the derivative of the 
dilaton field, as in the interpretation proposed in \cite{SS}.

The purpose
of this Letter is to present another geometrical formulation based in a
metric-compatible connection, valid for any space-time dimension.
The lagrangian will be the scalar of curvature of a
 metric-compatible connection,  and the interpretation of
the $H_{\alpha\mu\nu}$ field is compatible
with the previous results. We remember that the usual interpretation
of the torsion tensor $S_{\alpha\mu\nu}$
as being proportional to the totally anti-symmetrical
tensor $H_{\alpha\mu\nu}$ determines only the traceless part of 
$S_{\alpha\mu\nu}$. In our proposal, the dilaton field will determine
the trace of the torsion tensor and the volume-element. 
This is the key point,
the action principle will be formulated by using a non-parallel 
volume-element. Non-standard
 volume-elements have appeared sometimes in the literature,
as for example in the characterization of half-flats solutions of
 Einstein equations\cite{ASH}
and in the description of field theory on Riemann-Cartan spacetimes\cite{SAA}.
The proposed formulation has two advantages in comparison with the 
proposed in \cite{DT}. First, the space-time geometry is more simpler because
it involves a metric-compatible connection, and second, and more important,
our geometrical formulation applies also in the presence of matter fields,
elucidating the dilaton peculiar way of coupling.

We recall that the gravitational interaction involving
 the dilaton and the $B_{\mu\nu}$ fields is given by the equations
\begin{eqnarray}
\label{beta}
\beta^\Phi &=& 4D_\mu\Phi D^\mu\Phi - 4D^2\Phi 
-R + \frac{1}{12}H_{\alpha\beta\gamma}H^{\alpha\beta\gamma}=0, \nonumber \\
\beta^g_{\mu\nu} &=& R_{\mu\nu} 
- \frac{1}{4}H_\mu^{\ \lambda\rho}H_{\nu\lambda\rho} + 2D_\mu D_\nu\Phi=0, \\
\beta^B_{\mu\nu} &=& D_\lambda H^\lambda_{\ \mu\nu} 
- 2\left(D_\lambda\Phi \right)H^\lambda_{\ \mu\nu}=0, \nonumber
\end{eqnarray}
where $R_{\mu\nu}$, $R$, and $D_\mu$ are respectively the Ricci tensor,
the scalar of curvature, and the covariant derivative in the 
26-dimensional background 
manifold of metric $g_{\mu\nu}$, 
all these quantities being calculated by using the
Levi-Civita connection.  
The following conventions are adopted: 
${\rm sign}(g_{\mu\nu})=(+,-,-...)$, 
$R_{\alpha\nu\mu}^{\ \ \ \beta} = \partial_\alpha\Gamma_{\nu\mu}^\beta
+ \Gamma_{\alpha\rho}^\beta\Gamma_{\nu\mu}^\rho - 
(\alpha\leftrightarrow\nu)$,
$R_{\nu\mu}=R_{\alpha\nu\mu}^{\ \ \ \alpha}$, and
$g = |\det g_{\mu\nu}|$.
One can check that the equations (\ref{beta}) follow from the minimization 
of the action:
\begin{equation}
\label{action}
S = -\int d^{N}x\sqrt{g}e^{-2\Phi}\left( 
R + 4\partial_\mu\Phi \partial^\mu\Phi - \frac{1}{12}H_{\alpha\beta\gamma}
H^{\alpha\beta\gamma} 
\right).
\end{equation}
In spite of dilaton gravity is restricted by its origin in string theory
to space-times of dimension $N=26$, one can define stringy inspired models by
assuming that (\ref{action}) is valid also for other dimensions $N> 2$.
As to the  two dimensional case,
the totally anti-symmetrical third-rank tensor $H_{\alpha\beta\gamma}$
vanishes identically, and we have the following action for the two dimensional
dilaton gravity
\begin{equation}
\label{action2}
S = -\int d^{2}x\sqrt{g}e^{-2\Phi}\left( 
R + 4\partial_\mu\Phi \partial^\mu\Phi \right).
\end{equation}
We will discuss latter the incorporation of matter fields in 
(\ref{action}) and (\ref{action2}).

We wish to expound how the actions (\ref{action}) and (\ref{action2}) can be
understood as Hilbert-Einstein actions in a manifold endowed with a
metric-compatible connection and with a special volume-element.
To this end, let us introduce briefly some results on Riemann Cartan (RC)
manifolds. A RC manifold is a $N$-dimensional differentiable 
manifold endowed with a metric tensor $g_{\alpha\beta}(x)$ and with a 
metric-compatible connection $\Gamma_{\alpha\beta}^\mu$, which is 
non-symmetrical in its lower indices. From the anti-symmetric part of the
connection one can define the torsion tensor
$S_{\alpha\beta}^{\ \ \gamma} = \frac{1}{2}
\left(\Gamma_{\alpha\beta}^\gamma-\Gamma_{\beta\alpha}^\gamma \right).$
The metric-compatible connection, that is used to define the covariant
derivative ${\cal D}_\mu$, can be written as 
$\Gamma_{\alpha\beta}^\gamma = \left\{_{\alpha\beta}^\gamma \right\} 
- K_{\alpha\beta}^{\ \ \gamma},$
where $\left\{_{\alpha\beta}^\gamma \right\}$ are
 the usual Christoffel symbols, and 
$K_{\alpha\beta}^{\ \ \gamma}$ is the
contorsion tensor, which is given in terms of the torsion tensor by 
\begin{equation}
K_{\alpha\beta\gamma} = - S_{\alpha\beta\gamma} 
+ S_{\beta\gamma \alpha} - S_{\gamma \alpha\beta}.
\label{contorsion}
\end{equation}
The contorsion tensor (\ref{contorsion}) can be covariantly 
split in a traceless part and in a trace
\begin{equation}
K_{\alpha\beta\gamma} = \tilde{K}_{\alpha\beta\gamma} - 
\frac{2}{N-1}\left( 
g_{\alpha\gamma} S_\beta  - g_{\alpha\beta} S_\gamma
\right),
\label{decomposit}
\end{equation}
where $\tilde{K}_{\alpha\beta}^{\ \ \alpha}=0$
 and $S_\beta$ is 
the trace of the torsion tensor, $S_\beta = S^{\ \ \alpha}_{\alpha\beta}$.
The RC curvature tensor, calculated by using the 
connection $\Gamma_{\alpha\beta}^\mu$, 
according to our conventions is given by
\begin{equation}
{\cal R}_{\alpha\nu\mu}^{\ \ \ \ \beta} = \partial_\alpha \Gamma_{\nu\mu}^\beta
- \partial_\nu \Gamma_{\alpha\mu}^\beta 
+ \Gamma_{\alpha\rho}^\beta \Gamma_{\nu\mu}^\rho 
- \Gamma_{\nu\rho}^\beta \Gamma_{\alpha\mu}^\rho ,
\end{equation}
and after some manipulations we get the following expression for the 
scalar of curvature, valid for $N>2$,
\begin{equation} 
{\cal R} = g^{\mu\nu} {\cal R}_{\alpha\mu\nu}^{\ \ \ \ \alpha} =
R - 4{\cal D}_\mu S^\mu + \frac{4N}{N-1}S_\mu S^\mu - 
\tilde{K}_{\nu\rho\alpha} \tilde{K}^{\alpha\nu\rho},
\label{scurv}
\end{equation}
where $R$ is the Riemannian scalar of curvature, calculated from the
Christoffel symbols. For $N=2$ the RC scalar of curvature is given simply by
\begin{equation} 
{\cal R} = R - 4{D}_\mu S^\mu .
\label{scurv2}
\end{equation}

A volume-element in a differentiable orientable 
$N$-dimensional manifold $\cal M$ is an
$N$-form that does not vanish anywhere. If an $N$-form $\Omega$ defines
a volume-element, $f\Omega$ also does it if $f>0$. We see that the
volume-forms compatible with the orientation of $\cal M$ form a equivalence
class, $\{\Omega\}$. Assuming that $\cal M$ is endowed with a metric, the
local expression for an element of $\{\Omega\}$ is given by
\begin{equation}
\label{vol}
d^N\mu = f\sqrt{g} d^Nx,
\end{equation}
with $f>0$.
In the case that $\cal M$ has a metric, we have a privileged volume-element,
that one given by
$\Omega = 1^*$, which corresponds to the choice $f=1$ in (\ref{vol}). If
the manifold is endowed with a linear 
connection, another $N$-form $\Omega$ can be
singled out with the criterion of parallelism of the volume-form. We
know that in general cases they do not coincide\cite{SAA}. 
We will consider  for convenience  the one-parameter 
family of volume-elements
\begin{equation}
\label{vol1}
d^N\mu(\alpha) = e^{\alpha\Theta}\sqrt{g} d^Nx,
\end{equation}
with $\alpha$ a real parameter and $\Theta$ a scalar field.

As it was already said,  we will try to identify the trace of the
torsion tensor as a derivative of the dilaton field. To this end, let
us consider the anzats
\begin{equation}
\label{cond}
S_\beta(x) = \partial_\beta \Theta(x).
\end{equation}
We can now construct a Hilbert-Einstein action in a RC manifold 
using the volume (\ref{vol1})
and the condition (\ref{cond}). For the case $N>2$ we obtain
\begin{eqnarray}
\label{vaction}
S &=& -\int  \,d^N\mu(\alpha){\cal R}    \\
&=&-\int  \,d^Nx  \sqrt{g} e^{\alpha\Theta} \left( 
R + 4\left(\frac{N}{N-1}+\alpha -2 \right) 
\partial_\mu\Theta \partial^\mu \Theta 
- \tilde{K}_{\nu\rho\alpha} \tilde{K}^{\alpha\nu\rho} 
\right) + {\rm surf. \ terms}. \nonumber
\end{eqnarray}
We have that (\ref{vaction}) is identical to (\ref{action}) if one 
assumes that 
\begin{eqnarray}
\label{vactionx}
\Theta &=& -\frac{2}{\alpha} \Phi, \\ 
\tilde{K}_{\alpha\beta\gamma} &=& \frac{\sqrt{3}}{6}{H}_{\alpha\beta\gamma}, 
\nonumber \\
\alpha &=& \left\{ 
\begin{array}{l}
2 + 2/\sqrt{N-1} {\rm \ or} \\ 
2 - 2/\sqrt{N-1}
\end{array}
\right. \nonumber 
\end{eqnarray}

As to the two-dimensional case, the action will be
\begin{equation}
\label{vaction2}
S = -\int  \,d^2x  \sqrt{g} e^{\alpha \Theta}\left( 
R + 4\alpha \partial_\mu\Theta \partial^\mu \Theta  
\right),
\end{equation}
and to get (\ref{action2}) one needs 
$\Theta = -\frac{2}{\alpha} \Phi$, and 
$\alpha = 4$. The two dimensional case can be contrasted also with
the ref. \cite{DT2}, where the two-dimensional dilaton gravity is 
interpreted by means of a symmetrical non-Riemannian connection.

The simplest way to introduce matter fields into the discussion is
to add to the Hilbert-Einstein lagrangian a minimally coupled term 
describing their dynamics. The total lagrangian would be 
${\cal L = R + L}_{\rm matt}$. One can check that the action
$S = -\int d^N\mu(\alpha)\, {\cal L}$
with the assumptions (\ref{vactionx}) describes precisely the interaction
of the dilaton field with matter. Similar arguments can be applied to the
introduction of a cosmological constant.

As the conclusion, we summarize that the action (\ref{action}) is
a Hilbert-Einstein action for the metric-compatible 
connection and  volume-element given respectively in terms of the
dilaton and the $B_{\mu\nu}$ fields by
\begin{eqnarray}
\label{oo}
\Gamma_{\alpha\beta}^\gamma &=& \left\{_{\alpha\beta}^\gamma \right\} 
- \sqrt{3}/6H_{\alpha\beta}^{\ \ \gamma} - \frac{2}{N-1 + \sqrt{N-1}}\left( 
\delta^\gamma_\alpha \partial_\beta\Phi - g_{\alpha\beta}g^{\gamma\nu}
\partial_\nu\Phi
\right), \nonumber{\rm\ or} \\
&=& \left\{_{\alpha\beta}^\gamma \right\} 
- \sqrt{3}/6H_{\alpha\beta}^{\ \ \gamma} - \frac{2}{N-1 - \sqrt{N-1}}\left( 
\delta^\gamma_\alpha \partial_\beta\Phi - g_{\alpha\beta}g^{\gamma\nu}
\partial_\nu\Phi
\right), \nonumber \\
d^N\mu &=& d^N \sqrt{g} e^{-2\Phi},
\end{eqnarray}
for space-time dimensions greater than two. The two possible
connections correspond to the two values of $\alpha$ in (\ref{vactionx}).
The two-dimensional 
dilaton gravity can be obtained from a Hilbert-Einstein action constructed
from the following connection and volume-element,
\begin{eqnarray}
\Gamma_{\alpha\beta}^\gamma &=& \left\{_{\alpha\beta}^\gamma \right\} 
 - \left( 
\delta^\gamma_\alpha \partial_\beta\Phi - g_{\alpha\beta}g^{\gamma\nu}
\partial_\nu\Phi
\right), \nonumber \\
d^2\mu &=& d^2 \sqrt{g} e^{-2\Phi}.
\end{eqnarray}

 The author wishes
to thank CNPq for the financial support.

\end{document}